\documentclass[a4paper,11pt]{article}
\usepackage{jheppub}

\usepackage{graphicx} 
\usepackage{amssymb}
\usepackage{amsmath}
\usepackage{dsfont}
\usepackage{xcolor}
\usepackage{braket}
\usepackage[TS1,T1]{fontenc}
\usepackage{textcomp}

\title{The Triplet Perturbation of the Symmetric Orbifold}

\author{Matthias R.\ Gaberdiel and Ian Le Meur}

\affiliation{Institut f\"ur Theoretische Physik, ETH Zurich, \\
	CH-8093 Z\"urich, Switzerland}
\emailAdd{gaberdiel@itp.phys.ethz.ch,ilemeur@student.ethz.ch}

\abstract{The perturbation of the symmetric orbifold of $\mathbb{T}^4$ under the triplet of exactly marginal operators from the $2$-cycle twisted sector is studied in perturbation theory. We show that the structure of the triplet perturbation is very similar to that of the previously studied singlet perturbation, and in particular, that the theory remains also integrable in this case. Furthermore, using the various symmetries of the problem, we identify the dual supergravity interpretation of these deformations.}

\begin{document}

\numberwithin{equation}{section}
\maketitle

\newcommand{\wt}[1]{\tilde{#1}}
\newcommand{\mrm}[1]{\mathrm{#1}}
\newcommand{\mcl}[1]{\mathcal{#1}}
\newcommand{\mbb}[1]{\mathbb{#1}}
\newcommand{\mbf}[1]{\mathbf{#1}}
\newcommand{\mfr}[1]{\mathfrak{#1}}
\newcommand{\mscr}[1]{\mathscr{#1}}
\newcommand{\msf}[1]{\mathsf{#1}}
\newcommand{\mpzc}[1]{\mathpzc{#1}}
\newcommand{\bs}[1]{\boldsymbol{#1}}
\renewcommand{\d}{\delta}
\newcommand{\e}{\epsilon}
\renewcommand{\a}{\alpha}
\renewcommand{\b}{\beta}
\renewcommand{\c}{\gamma}

\newcommand{\wh}[1]{\widehat{#1}}
\newcommand{\ol}[1]{\bar{#1}}
\newcommand{\cint}[1]{\underset{#1}{\oint}}
\newcommand{\llangle}{\left\langle}
\newcommand{\rrangle}{\right\rangle}

\newcommand{\be}{\begin{equation}}
\newcommand{\ee}{\end{equation}}

\newcommand{\ba}{\begin{align}}

\section{Introduction}

The AdS/CFT dual for strings on ${\rm AdS}_3\times {\rm S}^3 \times \mathbb{T}^4$ is an  ${\cal N}=(4,4)$ 2d CFT that lies on the same moduli space as the symmetric orbifold of $\mathbb{T}^4$ \cite{Maldacena:1997re}. This low-dimensional case provides a very useful toy example since, unlike the case of strings on ${\rm AdS}_5$, a solvable worldsheet theory for strings on ${\rm AdS}_3$ is available \cite{Maldacena:2000hw,Berkovits:1999im}. While there was initially some confusion about which AdS background is precisely dual to the symmetric orbifold itself, see e.g.\ \cite{David:2002wn} for a review, a few years ago it was shown that it is the background with minimal ($k=1$) NS-NS flux \cite{Gaberdiel:2018rqv,Eberhardt:2018ouy,Eberhardt:2019ywk}. This has changed some of the old beliefs regarding the moduli space of string backgrounds, and some aspects of this were recently reconsidered in \cite{Aharony:2024fid}.

While the background with pure NS-NS flux is rather special, it is part of a $20$-dimensional moduli space, and one can consider the deformations of the theory in this moduli space. Many of the above moduli are in a sense uninteresting since they just describe the modifications of the torus, but there are $4$ `interesting' moduli that come from the $2$-cycle twisted sector of the symmetric orbifold and deform the theory away from the orbifold point. These $4$ moduli organise themselves as a singlet and a triplet with respect to an $\mathfrak{su}(2)$ symmetry --- the details of this will be described in Section~\ref{sec:symmetries} below --- and most attention so far has been focused on the singlet perturbation which is believed to be dual to switching on the R-R scalar in the dual AdS background, see e.g.\ \cite{Aharony:2024fid}.
\smallskip

Just like its higher-dimensional cousin, string theory on ${\rm AdS}_3$ is believed to be integrable \cite{Babichenko:2009dk}, and much evidence for this, mostly from the string theory perspective, has been found over the years \cite{Hoare:2013lja,Borsato:2013qpa,Lloyd:2014bsa}.\footnote{Mostly the deformation by the R-R scalar has been studied, but the case of the self-dual R-R 2-form was also analysed in \cite{OhlssonSax:2018hgc}.} On the other hand, an understanding of integrability in the dual symmetric orbifold CFT has remained somewhat elusive \cite{Lunin:2002fw,Gomis:2002qi,Gava:2002xb,David:2008yk,Pakman:2009mi}, and only recently
a first principle derivation for the case of the singlet perturbation was found in \cite{Gaberdiel:2023lco}, see also \cite{Gaberdiel:2024nge,Gaberdiel:2024dfw} for subsequent work. Among other things, this analysis provided direct evidence for the fact that the singlet perturbation corresponds to switching on the R-R scalar in the dual string theory. It is the aim of this paper to generalise these results to the triplet deformation. We will show that the perturbation by the triplet field leads to a very similar structure, and give strong evidence that it is dual to switching on the self-dual R-R 2-form in the dual string theory. 
\medskip

The paper is organised as follows. In Section~\ref{sec: symmetric orbifold} we introduce our conventions and review the calculation of \cite{Gaberdiel:2023lco}; we also give a careful explanation of the various symmetries of the problem. The perturbation analysis for the triplet perturbation is performed in Section~\ref{tripletpert}. Section~\ref{sec: supergravity} discusses the dual supergravity description, and provides evidence for the fact that the triplet perturbation in the symmetric orbifold is dual to switching on the self-dual R-R 2-form in the dual string theory. Our conclusions are contained in Section~\ref{sec: conclusion}, and there is an appendix, where we review some aspects of the symmetric orbifold theory.

\section{The singlet perturbation of the symmetric product orbifold}\label{sec: symmetric orbifold}

In this section we review the perturbation analysis of \cite{Gaberdiel:2023lco}, and explain how this will be generalised in the present paper. We begin by fixing our notation and reviewing basic facts about the conformal field theory with target space $\mbb{T}^4$ and its symmetric orbifold; we will mostly follow the conventions of \cite{Gaberdiel:2023lco}, see also \cite{David:2002wn} and \cite{Lunin:2000yv,Pakman:2009zz,Dei:2019iym} for useful references concerning the symmetric orbifold.

\subsection{The $\mbb{T}^4$ CFT and its symmetries}\label{sec:symmetries}

The conformal field theory with target space $\mbb{T}^4$ is an $\mcl{N}=(4,4)$ superconformal field theory that is generated by four free  bosons and fermions. We denote the left-moving four bosons\footnote{The bosons $X^{i}$ themselves are neither purely left- or right-moving: the `left-moving bosons' we discuss here are the associated chiral currents $\alpha^i = \partial  X^i$, while the `right-moving bosons' will be $\tilde{\alpha}^i = \bar\partial X^i$, and similarly for $\bar{X}^i$.} and four fermions of the $\mbb{T}^4$ by $\a^i$, $\bar{\a}^i$, $i=1,2$ and $\psi^\pm$, $\bar{\psi}^\pm$, respectively. They satisfy the OPE relations
\be
\bar{\a}^i(x)\a^j(y)\sim \frac{\epsilon^{ij}}{(x-y)^2}\ ,\qquad \bar{\psi}^\pm(x)\psi^\mp(y)\sim \frac{\pm 1}{x-y} \ ,
\ee
where $\epsilon^{ij}$ is antisymmetric with $\epsilon^{12}=+1$. We will always use the convention that right-movers are denoted by a tilde. 
The ${\cal N}=4$ superconformal generators  can be expressed in terms of bilinears of these free fields, and explicit expressions are given in Appendix~\ref{app:algebra}.

For the following it will be useful to describe in some detail the various symmetries of the $\mathbb{T}^4$ theory; these symmetries will also be inherited by the symmetric orbifold, and they play an important role in our analysis. First of all, the four left-moving free fermions give rise to an $\hat{\mathfrak{so}}(4)_1 $ affine symmetry, which we can write as 
\be
\hat{\mathfrak{so}}(4)_1 \cong \hat{\mathfrak{su}}(2)_1^{(1)}  \oplus \hat{\mathfrak{su}}(2)_1^{(2)}  \ , 
\ee 
using the well-known isomorphism of $\mathfrak{so}(4) \cong \mathfrak{su}(2) \oplus \mathfrak{su}(2)$. We shall work with the convention that the first factor is the R-symmetry $\mathfrak{su}(2)$ that is part of the ${\cal N}=4$ superconformal algebra, 
\be
\hat{\mathfrak{su}}(2)_1^{(1)}  \equiv \hat{\mathfrak{su}}(2)_{\rm R}: \qquad \hbox{generated by $K^a_n$ with $a=3,\pm$ and $n\in \mathbb{Z}$,}
\ee
where the $K^a_n$ generators are defined in Appendix~\ref{app:algebra}. We shall often only be interested in the zero-modes of these affine algebras, and we shall denote them by $\mathfrak{su}(2)_{\rm R}$ for $\hat{\mathfrak{su}}(2)_{\rm R}$, and $\mathfrak{su}(2)_{\rm F}^{(2)}$ for 
$\hat{\mathfrak{su}}(2)_1^{(2)}$, respectively. The discussion for the right-moving generators is analogous, and we shall distinguish the corresponding algebras by a tilde. 
\smallskip

For the free bosons the situation is a bit more complicated since they do not give rise to an affine symmetry. However, if we ignore the underlying momentum and winding lattice, the free fields themselves sit in the vector representation of a global $\mathfrak{so}(4)_{\rm B}$ algebra that rotates the different bosonic modes into one another. This global $\mathfrak{so}(4)_{\rm B}$ algebra acts simultaneously on left- and right-moving bosonic oscillator modes,\footnote{Since the left- and right-moving modes commute with one another, one can, however, also consistently define independent left- and right-moving actions on these bosonic modes, and we shall sometimes do so, see e.g.\ the discussion below eq.~(\ref{3.1}).\label{foot2}} and we can write it again in terms of two $\mathfrak{su}(2)$ algebras
\be
\mathfrak{so}(4)_{\rm B} \cong \mathfrak{su}(2)_{\rm B}^{(1)} \oplus  \mathfrak{su}(2)_{\rm B}^{(2)}\ . 
\ee
For the following, it will be convenient to consider the diagonal subalgebra of 
\be
\mathfrak{su}(2)_{[2]} \equiv \Bigl( \mathfrak{su}(2)_{\rm F}^{(2)} \oplus \tilde{\mathfrak{su}}(2)_{\rm F}^{(2)} \oplus \mathfrak{su}(2)_{\rm B}^{(2)} \Bigr)_{\rm diagonal} \ , 
\ee
that will leave the ${\cal N}=(4,4)$ generators invariant. This $\mathfrak{su}(2)$ then commutes with the two R-symmetries, as well as $\mathfrak{su}(2)_{[1]} \equiv \mathfrak{su}(2)_{\rm B}^{(1)}$. Altogether, we can therefore organise all states in representations of the commuting $\mathfrak{su}(2)$ algebras
\be
\mathfrak{su}(2)_{\rm R} \oplus \tilde{\mathfrak{su}}(2)_{\rm R} \oplus \mathfrak{su}(2)_{[1]} \oplus \mathfrak{su}(2)_{[2]}  \ .
\ee
With these conventions, the left-moving free fermions transform in the $({\bf 2},{\bf 2})$ with respect to $\mathfrak{su}(2)_{\rm R}\oplus \mathfrak{su}(2)_{[2]}$, i.e.\ we can write them as 
\begin{equation}\label{genferm}
    \bs{{\Psi}}_{\a i}=\begin{pmatrix}
        {\psi}^+ & - {\bar{\psi}}^+ \\
        {\psi}^- & - {\bar{\psi}}^-
    \end{pmatrix}\ , \qquad  \qquad \begin{array}{ll}
     \hbox{left-action:}  \ \ &  \mathfrak{su}(2)_{\rm R} \\
     \hbox{right-action:} \ \ &  \mathfrak{su}(2)_{[2]}    \ ,  \end{array} 
\end{equation}
where $\mathfrak{su}(2)_{\rm R}$ acts from the left, while $\mathfrak{su}(2)_{[2]}$  acts from the right. We will always use Latin letters from the middle of the alphabet to denote an index with respect to $\mfr{su}(2)_{[2]}$, and Greek letters to denote an index with respect to $\mathfrak{su}(2)_{\rm R}$ and $\tilde{\mathfrak{su}}(2)_{\rm R}$. The right-moving fermions transform similarly, except that $\mathfrak{su}(2)_{\rm R}$ is replaced by $\tilde{\mathfrak{su}}(2)_{\rm R}$. For the left- and right-moving bosons we have instead
\begin{equation}\label{genbos}
    \bs{{\mcl{B}}}_{i{a}}=\begin{pmatrix}
        \a^2 & \a^1 \\
        {\bar{\a}}^2 & {\bar{\a}}^1
    \end{pmatrix}\ , \qquad  \qquad \qquad \begin{array}{ll}
     \hbox{left-action:}  \ \ & \mathfrak{su}(2)_{[2]} \\
     \hbox{right-action:} \ \ & \mathfrak{su}(2)_{[1]} \ ,
     \end{array} 
\end{equation}
where now $\mathfrak{su}(2)_{[2]}$ acts from the left and $\mathfrak{su}(2)_{[1]}$  from the right, and we use Latin letters from the beginning of the alphabet to denote an index with respect to $\mfr{su}(2)_{[1]}$. With these conventions, the ${\cal N}=4$ supercurrents are the $\mathfrak{su}(2)_{[2]}$ invariant bilinears 
\begin{equation}
    \bs{G}_{\a a}=\bs{{\Psi}}_{\a i}\,\e_{ij}\,\bs{{\mcl{B}}}_{j{a}}\ ,\qquad \bs{\wt G}_{\a a}=\bs{\wt {\Psi}}_{\a i}\,\e_{ij}\,\bs{\wt {\mcl{B}}}_{j{a}}\ \  ,
\end{equation}
and they satisfy the algebra relations of eq.~(\ref{N4rel}). The relation to the conventions used in  \cite{Gaberdiel:2023lco} is  
\begin{equation}\label{GtildeG}
    \bs{G}_{\a a}=\begin{pmatrix}
        G^+  & - G^{'+} \\
        G^{'-}  &  G^{-}
    \end{pmatrix}\ , \qquad   \bs{\tilde{G}}_{\a a}=\begin{pmatrix} 
        \tilde{G}^+  & - \tilde{G}^{'+} \\
        \tilde{G}^{'-}  &  \tilde{G}^{-}
    \end{pmatrix}\ .
\end{equation}
We note that the left-moving supercurrents $\bs{G}_{\a a}$ transform in the $({\bf 2},{\bf 2})$ with respect to $\mathfrak{su}(2)_{\rm R}$ and $\mathfrak{su}(2)_{[1]}$, while their right-moving analogues $\bs{\tilde{G}}_{\a a}$ transform in the $({\bf 2},{\bf 2})$ with respect to $\tilde{\mathfrak{su}}(2)_{\rm R}$ and $\mathfrak{su}(2)_{[1]}$.

\subsection{The symmetric orbifold of $\mathbb{T}^4$}

The symmetric orbifold of $\mbb{T}^4$ consists of $N$ copies of the $\mbb{T}^4$ theory, which we orbifold by the symmetric group $S_N$, acting on the $N$ copies.\footnote{In particular, the $S_N$ action commutes with the diagonal sum of the above $\mathfrak{su}(2)$ actions, where each $\mathbb{T}^4$ is transformed in the same manner.} In addition to the untwisted sector which just consists of the orbifold invariant states of the $\mbb{T}^{4N}$ theory, we have twisted sectors corresponding to the conjugacy classes of the permutation group $S_N$. Since we are primarily interested in the single particle excitations (that are dual to the excitations of a single worldsheet in the corresponding AdS string description), we restrict ourselves to the twisted sectors associated to single-cycle permutations, and we denote the length of the corresponding cycle by $w$. In the $w$-cycle twisted sector, the free fields have $\frac{1}{w}$-fractional modes, and the whole $w$-cycle twisted sector is generated by the action of orbifold-invariant combinations of these fractional modes on some suitable reference state. It is convenient to take this reference state to be the BPS state $|{\rm BPS}\rangle_w$ with conformal charge $h$ and $\mathfrak{su}(2)$ spin $j$, 
\be\label{BPS}
|{\rm BPS}\rangle_w: \qquad h = j =m = \frac{w+1}{2} \ , \qquad \tilde{h} = \tilde{\jmath} = \tilde{m} = \frac{w+1}{2} \ . 
\ee
(Here, as below, $m$ denotes the eigenvalue of the Cartan generator of the $\mathfrak{su}(2)_{\rm R}$ algebra $K^3_0$, and similarly for $\tilde{m}$.) The orbifold invariance condition is then that the difference of left- minus right-moving modes is an integer for bosons and a half-integer for fermions, see \cite{Lunin:2000yv,Pakman:2009zz,Dei:2019iym,Gaberdiel:2023lco} for further details.

\subsection{The perturbation of the symmetric orbifold}

Following \cite{Lunin:2002fw,Gomis:2002qi,Gava:2002xb} we are interested in perturbing the symmetric orbifold by an exactly mar\-ginal operator that arises from the $w=2$ cycle twisted sector. 
The ground states of the $w=2$ cycle twisted sector consist of four states with $h = \tilde{h} = \tfrac{1}{2}$ that transform in the $({\bf 2},{\bf 2})$ with respect to $\mathfrak{su}(2)_{\rm R} \oplus \tilde{\mathfrak{su}}(2)_{\rm R}$, but are singlets with respect to $\mathfrak{su}(2)_{[1]}$ and $\mathfrak{su}(2)_{[2]}$. The exactly marginal operators are the $h=\tilde{h}=1$ supercharge descendants of these states that transform as singlets with respect to $\mathfrak{su}(2)_{\rm R} \oplus \tilde{\mathfrak{su}}(2)_{\rm R}$ (so as not to break the ${\cal N}=(4,4)$ superconformal symmetry). Since the supercharges transform as $({\bf 2},{\bf 2})$ with respect to $\mathfrak{su}(2)_{\rm R}$ (resp.\ $\tilde{\mathfrak{su}}(2)_{\rm R}$) and $\mathfrak{su}(2)_{[1]}$, there are four exactly marginal operators that transform as 
\be
{\bf 2} \otimes {\bf 2} = {\bf 1} \oplus {\bf 3} \qquad \hbox{with respect to $\mathfrak{su}(2)_{[1]}$.}
\ee
In fact, using various null-vector relations, see e.g.\ \cite[eqs.~(2.1) and (2.2)]{Fiset:2022erp}, we can write these four states as 
\begin{subequations}\label{eq: four perturbations}
\ba
    \Phi_{\text{singlet}}& =\frac{i}{\sqrt{2}}\left(G^-_{-\frac{1}{2}}\wt{G}'^{-}_{-\frac{1}{2}}-G'^{-}_{-\frac{1}{2}}\wt{G}^-_{-\frac{1}{2}}\right)\lvert \text{BPS}_- \rangle_2\ , \\
    \Phi_{x}& =\frac{i}{\sqrt{2}}\left(G'^-_{-\frac{1}{2}}\wt{G}'^{-}_{-\frac{1}{2}}-G^{-}_{-\frac{1}{2}}\wt{G}^{-}_{-\frac{1}{2}}\right)\lvert \text{BPS}_- \rangle_2\ , \\
    \Phi_{y}& =-\frac{1}{\sqrt{2}}\left(G^{-}_{-\frac{1}{2}}\wt{G}^{-}_{-\frac{1}{2}}+G'^-_{-\frac{1}{2}}\wt{G}'^{-}_{-\frac{1}{2}}\right)\lvert \text{BPS}_- \rangle_2\ , \\
    \Phi_{z}& =-\frac{i}{\sqrt{2}}\left(G^-_{-\frac{1}{2}}\wt{G}'^-_{-\frac{1}{2}}+G'^-_{-\frac{1}{2}}\wt{G}^-_{-\frac{1}{2}}\right)\lvert \text{BPS}_- \rangle_2\ . 
\end{align}
\end{subequations}
Here $\lvert \text{BPS}_- \rangle_2$ is the BPS ground state with charges\footnote{Relative to the BPS state $|{\rm BPS}\rangle_w$ of eq.~(\ref{BPS}), which is the `upper' BPS state in the $w$-cycle twisted sector, $\lvert \text{BPS}_- \rangle_2$ is the `lower' BPS state in the $w=2$ cycle twisted sector, see Appendix~\ref{app:states} for a more detailed exposition.}
\be
\lvert \text{BPS}_- \rangle_2: \qquad h = \tilde{h} = \tfrac{1}{2} \ , \quad m = \tilde{m} = \tfrac{1}{2} \ , 
\ee
where $m$ and $\tilde{m}$ are again the eigenvalues under the Cartan generators $K^3_0$ and $\tilde{K}^3_0$ of the $R$-symmetry algebras $\mathfrak{su}(2)$ and $\tilde{\mathfrak{su}}(2)$,  respectively. Furthemore, $\Phi_{\text{singlet}}$ is the singlet with respect to the $\mathfrak{su}(2)_{[1]}$ action, while the other three perturbing fields $\Phi_{x,y,z}$ transform as a triplet, i.e.\ as a ${\bf 3}$ under $\mathfrak{su}(2)_{[1]}$.

\subsection{The singlet perturbation}\label{singletpert}

The analysis for the singlet perturbation was done in detail in \cite{Gaberdiel:2023lco}, see also \cite{Gaberdiel:2024nge,Gaberdiel:2024dfw} for further developments. The result is best described in terms of the `magnons' that parametrise the fractional modes in the $w$-cycle twisted sector, 
\be\label{modes}
\begin{aligned}
& \psi^-( \tfrac{n}{w}) \equiv  \psi^-_{-\frac{1}{2} + \frac{n}{w}} \ , \ \
\psi^+( \tfrac{n}{w}) \equiv \psi^+_{-\frac{3}{2} + \frac{n}{w}} \ ,\qquad
\a^i( \tfrac{n}{w}) \equiv \tfrac{1}{\sqrt{ 1-\frac{n}{w}}} \, \a^i_{-1 + \frac{n}{w}}  \ \ (i=1,2)  \ ,
\\
& \bar{\psi}^-( \tfrac{n}{w}) \equiv \bar{\psi}^-_{-\frac{1}{2} + \frac{n}{w}} \ , \ \
\bar{\psi}^+( \tfrac{n}{w}) \equiv \bar{\psi}^+_{-\frac{3}{2} + \frac{n}{w}} \ ,\qquad
\bar{\a}^i( \tfrac{n}{w}) \equiv \tfrac{1}{\sqrt{1-\frac{n}{w}}} \, \bar{\a}^i_{-1 + \frac{n}{w}}  \ \ (i=1,2) \ ,
\end{aligned}
\ee
and similarly for the right-moving modes. The anomalous conformal dimension can be most easily obtained by studying the behaviour of the left-moving magnons under the action of the global supersymmetry modes\footnote{A mnemonic is that the $S$'s are the $-\frac{1}{2}$ raising modes while the $Q$'s are the  $+\frac{1}{2}$ lowering modes. The subscripts $(1,2)$ refer to the unprimed and primed generators, respectively.}  
\be\label{superc}
\begin{aligned} 
& Q_1 \equiv G^-_{+\frac{1}{2}} \ , \qquad S_1 \equiv G^+_{-\frac{1}{2}} \ , \qquad
Q_2 \equiv {G}'^{-}_{+\frac{1}{2}} \ , \qquad S_2 \equiv {G}^{'+}_{-\frac{1}{2}} \\
& \tilde{Q}_1 \equiv \tilde{G}^-_{+\frac{1}{2}} \ , \qquad \tilde{S}_1 \equiv \tilde{G}^+_{-\frac{1}{2}} \ , \qquad
\tilde{Q}_2 \equiv \tilde{G}'^{-}_{+\frac{1}{2}} \ , \qquad \tilde{S}_2 \equiv \tilde{G}^{'+}_{-\frac{1}{2}} \ ,
\end{aligned}
\ee
for which the only non-vanishing anti-commutators are 
\be\label{anti-com}
\{ Q_1, S_1 \}  = \{ Q_2, S_2 \}  = {\cal C} \equiv (L_0 - K^3_0) \ , \qquad 
\{ \tilde{Q}_1, \tilde{S}_1 \}  = \{ \tilde{Q}_2, \tilde{S}_2 \}  = \tilde{\cal C} \equiv (\tilde{L}_0 - \tilde{K}^3_0) \ .
\ee
For example, it was shown in \cite{Gaberdiel:2023lco} that to first order in the singlet perturbation one finds 
\be\label{deformedaction}
\begin{array}{llll}
Q_1\, \a^2( \tfrac{n}{w}) \, |{\rm BPS}\rangle_w & = a^n_n\, \psi^-( \tfrac{n}{w}) \, |{\rm BPS}\rangle_w   \quad
& \wt{Q}_2\,  \a^2( \tfrac{n}{w}) \, |{\rm BPS}\rangle_w & =  0  \\[2pt]
Q_1\, \psi^-( \tfrac{n}{w})  \, |{\rm BPS}\rangle_w & = 0  \quad
& \wt{Q}_2\,  \psi^-( \tfrac{n}{w})  \, |{\rm BPS}\rangle_w & =  b^m_n \,
\a^2( \tfrac{m}{w-1}) \, {\cal Z}_- \, |{\rm BPS}\rangle_w  \\[2pt]
S_1\, \a^2( \tfrac{n}{w}) \, |{\rm BPS}\rangle_w & = 0   \quad
& \wt{S}_2\,  \a^2( \tfrac{n}{w}) \, |{\rm BPS}\rangle_w & =  c^m_n \,
\psi^-( \tfrac{m}{w+1}) \, {\cal Z}_+ \, |{\rm BPS}\rangle_w \\[2pt]
S_1\, \psi^-( \tfrac{n}{w}) \, |{\rm BPS}\rangle_w & = d^n_n \, \a^2( \tfrac{n}{w}) \, |{\rm BPS}\rangle_w  \quad
& \wt{S}_2\,  \psi^-( \tfrac{n}{w})  \, |{\rm BPS}\rangle_w & =  0 \ ,
\end{array}
\ee
where ${\cal Z}_\pm$ maps the reference BPS states of different twisted sectors into one another,
\be\label{Zdef}
{\cal Z}_\pm \, |{\rm BPS}\rangle_w = |{\rm BPS}\rangle_{w\pm 1} \ ,
\ee
and $a^n_n$, $b^m_n$, $c^m_n$ and $d^n_n$ are matrices that can be computed as correlators using the covering map method;\footnote{We shall only work to first order in the perturbation; then $a$ and $d$ are diagonal matrices (determined from the usual action of the left-moving supercharges on the left-moving modes), and only $b^m_n$ and $c^m_n$ are non-trivial.} for example, for the case of the $c$ matrix one finds 
\be
c^{m}_n  = g\, \underset{C(0)}{\oint}d\bar{x}\int d^2y\,  {}_{w+1}\!\langle {\rm BPS} |  \, \bigl(\psi^-(\tfrac{m}{w+1})\bigr)^\dagger\, \, \wt{G}'^+(\bar{x})\,\Phi_{\rm{singlet}}(y,\bar{y})\, \, {\a}^2(\tfrac{n}{w}) | {\rm BPS}\rangle_w \ , 
\label{3.27}
\ee
where $g$ is the coupling constant for the singlet perturbation, and the integral was computed explicitly in \cite{Gaberdiel:2023lco}. While the result at finite $w$ is quite complicated, it simplifies nicely in the limit where one takes the twist $w$ large, while fixing the `magnon momentum' $p = \frac{n}{w}$. The other cases could be determined similarly, and the final result can be summarised succinctly as\footnote{In the following we only give the action of the left-moving supercharges on the left- and right-moving magnons; the formula for the action of the right-moving supercharges is analogous.}
\begin{align}
[Q_1,\alpha^2(p)]&= \eta_p\,\psi^-(p)\ , & \{Q_1,\psi^+(p)\} &= -\eta_p\,\alpha^1(p)\ ,\nonumber\\
[Q_1,\tilde{\alpha}^2(p)] &= e^{-i\pi p}\frac{\eta_p}{x_p}\,\tilde{\psi}^+(p)\,\mathcal{Z}_-\ , & \{Q_1,\tilde{\psi}^-(p)\} &=  e^{-i\pi p}\frac{\eta_p}{x_p}\,\tilde{\alpha}^1(p)\,\mathcal{Z}_-\ ,\nonumber\\
\end{align}
\begin{align}
[Q_2,\alpha^1(p)]&= -\eta_p\,\psi^-(p)\ , & \{Q_2,\psi^+(p)\} &= -\eta_p\,\alpha^2(p)\ ,\nonumber\\
[Q_2,\tilde{\alpha}^1(p)] &= - e^{-i\pi p}\frac{\eta_p}{x_p}\,\tilde{\psi}^+(p)\,\mathcal{Z}_-\ , & \{Q_2,\tilde{\psi}^-(p)\} &=  e^{-i\pi p}\frac{\eta_p}{x_p}\,\tilde{\alpha}^2(p)\,\mathcal{Z}_-\ ,\nonumber\\
&&&\nonumber\\
[S_1,\alpha^1(p)]&= -\eta_p\,\psi^+(p)\ , & \{S_1,\psi^-(p)\} &= \eta_p\,\alpha^2(p)\ ,\nonumber\\
[S_1,\tilde{\alpha}^1(p)] &= e^{i\pi p}\frac{\eta_p}{x_p}\,\tilde{\psi}^-(p)\,\mathcal{Z}_+\ , & \{S_1,\tilde{\psi}^+(p)\} &= e^{i\pi p}\frac{\eta_p}{x_p}\,\tilde{\alpha}^2(p)\,\mathcal{Z}_+\ ,\nonumber\\
&&&\nonumber\\
[S_2,\alpha^2(p)]&= -\eta_p\,\psi^+(p)\ , & \{S_2,\psi^-(p)\} &= -\eta_p\,\alpha^1(p)\ ,\nonumber\\
[S_2,\tilde{\alpha}^2(p)] &= e^{i\pi p}\frac{\eta_p}{x_p}\,\tilde{\psi}^-(p)\,\mathcal{Z}_+\ , & \{S_2,\tilde{\psi}^+(p)\} &= -e^{i\pi p}\frac{\eta_p}{x_p}\,\tilde{\alpha}^1(p)\,\mathcal{Z}_+\ . \label{eq: full sym alg}
\end{align}
Here we have used the Zhukovski variables defined as 
\begin{align}\label{coeff-mom-dep}
\eta_p &=  \sqrt{\tfrac{1}{2}\bigl((1-p)+\epsilon(p)\bigr)}\ , \qquad
\frac{\eta_p}{x_p}e^{i\pi p} = \frac{g}{2i}\,\frac{e^{2\pi i p}-1}{\sqrt{\tfrac{1}{2}\bigl((1-p) + \epsilon(p) \bigr)}}\ ,
\end{align}
that satisfy the conditions
\begin{equation}
\eta_p^2 = g\,x_p\,\sin(\pi p)\ ,\qquad x_p-\frac{1}{x_p} = \frac{1-p}{g\,\sin(\pi p)}\ .
\end{equation}
Applying for example $\{ Q_1, S_1 \}  = \{ Q_2, S_2 \}  = {\cal C} \equiv (L_0 - K^3_0)$ to a right-moving magnon excitation, we then find 
\be\label{singletano}
\{ Q_1, S_1 \}   \, \tilde{\alpha}^i(\tilde{p}) | {\rm BPS}\rangle_w =  \left(\frac{\eta_{\wt p}}{x_{\wt p}}\right)^2 \, 
\tilde{\alpha}^i(\tilde{p}) | {\rm BPS}\rangle_w \ , 
\ee
and similarly for the other modes. As in \cite[Section~5.5]{Gaberdiel:2023lco} this then allows one to show that all magnon modes have the anomalous dispersion relation 
\be\label{disp}
\epsilon(p) = \sqrt{ (1-p)^2 + 4 \, g^2 \, \sin^2(\pi p)} \ . 
\ee

\section{The triplet perturbation}\label{tripletpert}

In this section we explain how the perturbation analysis of Section~\ref{singletpert} can be generalised to the case where the perturbing field is in the triplet representation. As it turns out, it is actually simpler to treat all four perturbations in (\ref{eq: four perturbations}) on an equal footing by defining 
\begin{equation}\label{3.1}
    \bs{\Phi}_{a\tilde{a}}=\frac{i}{\sqrt{2}}\begin{pmatrix}
        G_{-\frac{1}{2}}'^-\wt{G}_{-\frac{1}{2}}'^- & G_{-\frac{1}{2}}'^-\wt{G}_{-\frac{1}{2}}^- \\
        G_{-\frac{1}{2}}^-\wt{G}_{-\frac{1}{2}}'^- & G_{-\frac{1}{2}}^-\wt{G}_{-\frac{1}{2}}^-
    \end{pmatrix}|\text{BPS}_- \rangle_2\ . 
\end{equation}
Here we think of $a$ and $\tilde{a}$ as spinor indices (with values $\pm$) with respect to $\mfr{su}(2)_{[1]}$ and $\tilde{\mfr{su}}(2)_{[1]}$, respectively, i.e.\ we take $\mfr{su}(2)_{[1]}$ to act only on the left-moving bosonic modes, while $\tilde{\mfr{su}}(2)_{[1]}$ only rotates the right-moving bosonic modes.\footnote{While this is not an actual symmetry of the full theory, it is nevertheless useful to organise the perturbing fields in this manner, see Footnote~\ref{foot2} above.}  

We then repeat the above perturbation analysis, except that we replace $\Phi_{\rm singlet}$ by $\bs{\Phi}_{a\tilde{a}}$ in eq.~(\ref{3.27}), and similarly for the other matrix elements. 
It is natural to believe that the above commutation resp.\ anti-commutation relations transform covariantly with respect to the different $\mathfrak{su}(2)$ algebras, and this indeed turns out to be case. Explicitly, we find by a straightforward (but somewhat tedious) calculation\footnote{The different matrix elements that appear below can also be determined from those in \cite{Gaberdiel:2023lco} by using the $\mfr{su}(2)_{[1]}$ and $\tilde{\mfr{su}}(2)_{[1]}$ transformations of eq.~(\ref{3.1}).}
\begin{subequations}\label{3.2}
\begin{align}
    \relax[{\bs{G}}_{\a a}, \bs{\wt{\mcl{B}}}_{\wt{\imath}\wt{a}}(\wt{p})]_{\bs{\Phi}_{b\wt{b}}}&= -\e_{\a\b}\e_{ab}\e_{\wt{a}\wt{b}}e^{\a i\pi \wt{p}}\frac{\eta_{\wt{p}}}{x_{\wt{p}}}\bs{\wt{\Psi}}_{\b\wt{\imath}}(\wt{p})\mcl{Z}_{\a}\ , \\
    [\bs{\wt{G}}_{\a \wt{a}}, \bs{{\mcl{B}}}_{i{a}}({p})]_{\bs{\Phi}_{b\wt{b}}}&= \e_{\a\b}\e_{{a}b}\e_{\wt a\wt b}e^{-\a i\pi {p}}\frac{\eta_{{p}}}{x_{{p}}}\bs{{\Psi}}_{\b i}({p})\mcl{Z}_{\a}\ , \\
     \{\bs{G}_{\a a}, \bs{\wt{\Psi}}_{\b\wt{\imath}}(\wt p)\}_{\bs{\Phi}_{b\wt{b}}}&= \d_{\a\b}\e_{ab}e^{\a i\pi \wt p}\frac{\eta_{\wt p}}{x_{\wt p}}\bs{\wt{\mcl{B}}}_{\wt{\imath}\wt{b}}(\wt p)\mcl{Z}_{\a}\ ,\\
     \{\bs{\wt G}_{\a \wt a}, \bs{{\Psi}}_{\b i}( p)\}_{\bs{\Phi}_{b\wt{b}}}&= -\d_{\a\b}\e_{\wt a\wt b}e^{-\a i\pi  p}\frac{\eta_{ p}}{x_{p}}\bs{{\mcl{B}}}_{ ib}( p)\mcl{Z}_{\a}\ ,\\
    [\bs{G}_{\a a}, \bs{\mcl{B}}_{ib}(p)]&= -\e_{ab}\eta_p\bs{\Psi}_{\a i}(p)\ , \\
    [\bs{\wt G}_{\a \wt a}, \bs{\wt{\mcl{B}}}_{\wt{\imath}\wt b}(\wt p)]&= -\e_{\wt a\wt b}\eta_{\wt p}\bs{\wt \Psi}_{\a\wt{\imath}}(\wt p)\ , \\
    \{\bs{G}_{\a a}, \bs{\Psi}_{\b i}(p)\}&= \e_{\a\b}\eta_p\bs{\mcl{B}}_{ia}(p)\ , \\
    \{\bs{\wt G}_{\a \wt a}, \bs{\wt \Psi}_{\b\wt{\imath}}(\wt p)\}&= \e_{\a\b}\eta_{\wt p}\bs{\wt{\mcl{B}}}_{\wt{\imath}\wt a}(\wt p)\ ,
\end{align}
\end{subequations}
where $\mcl{B}_{i{a}}$ denote the bosonic generators of eq.~(\ref{genbos}), while $\bs{{\Psi}}_{\a i}$ are the fermions of eq.~(\ref{genferm}), and similarly for the right-movers. Furthermore, the supercharges $\bs{G}_{\a a}$ and $\tilde{\bs{G}}_{\a a}$ refer here to the corresponding $\pm \frac{1}{2}$ modes, 
\begin{equation}\label{GtildeGm}
    \bs{G}_{\a a}=\begin{pmatrix}
        G^+_{-\frac{1}{2}}  & - G^{'+}_{-\frac{1}{2}} \\
        G^{'-}_{+\frac{1}{2}}  &  G^{-}_{+\frac{1}{2}} 
    \end{pmatrix}\ , \qquad   \bs{\tilde{G}}_{\a a}=\begin{pmatrix} 
        \tilde{G}^+_{-\frac{1}{2}}  & - \tilde{G}^{'+}_{-\frac{1}{2}}  \\
        \tilde{G}^{'-}_{+\frac{1}{2}}   &  \tilde{G}^{-}_{+\frac{1}{2}} 
    \end{pmatrix}\ .
\end{equation}
Finally, the $\epsilon$ tensor is anti-symmetric with non-vanishing matrix elements $\e_{+-}=-\e_{-+}=1$, and 
$e^{\a i \pi p}\mcl{Z}_{\a}$ equals $e^{i \pi p}\mcl{Z}_{+}$ if $\a=+$ and $e^{-i \pi p}\mcl{Z}_{-}$ if $\a=-$, and similarly for $e^{-\a i \pi p}\mcl{Z}_{\a}$.

\subsection{Closure of the algebra}

Before proceeding it is useful to subject these results to various consistency checks. First we should confirm that also the general perturbation preserves the $\mcl{N}=(4,4)$ algebra on physical states. Recall that the physical (= orbifold invariant) states are those  multi-magnon states 
\begin{equation*}
    \ket{A_1(p_1),\cdot\cdot\cdot, A_l(p_l),\wt A_1(\wt p_1),\cdot\cdot\cdot, \wt A_m(\wt p_m)}\equiv\lim_{w\to \infty}A_1(p_1)\cdot\cdot\cdot A_l(p_l)\wt A_1(\wt p_1) \cdot\cdot\cdot \wt A_m(\wt p_m)\ket{\text{BPS}}_w
\end{equation*}
which satisfy
\begin{equation}
    \sum_{i=1}^lp_i-\sum_{j=1}^m\wt p_j\in\mbb{Z}\ . \label{eq: phys state cond}
\end{equation}
Here $A$ and $\wt A$ are  left- and right-moving magnons, respectively, and they can be either bosons or fermions. For the actual $\mcl{N}=(4,4)$ algebra, all anti-commutators $\{\bs{G}_{\a a},\bs{\wt G}_{\b\wt a}\}$ must vanish, but this is not manifest from the above formulae, and in fact, this will only turn out to be true on states satisfying the physical state condition eq.~(\ref{eq: phys state cond}). To check this we need to calculate 
\begin{equation}\label{3.6}
    \{\bs{G}_{\a a},\bs{\tilde G}_{\b\wt a}\}\ket{A_1(p_1),\cdot\cdot\cdot, A_l(p_l),\wt A_1(\wt p_1),\cdot\cdot\cdot, \wt A_m(\wt p_m)} \ . 
\end{equation}
Because of the fermionic statistics, the only terms that contribute are those where both $\bs{G}_{\a a}$ and $\bs{\tilde G}_{\b\wt a}$ act on the same magnon, and for the different cases we find 
\be
\begin{aligned}
    \{\bs{G}_{\a a},\bs{\wt G}_{\b\wt a}\}\bs{\mcl{B}}_{ic}(p)|_{\bs{\Phi}_{b\wt b}}&=\d_{\a\b}\e_{\wt a \wt b}e^{-\b i\pi p}\frac{\eta_p^2}{x_p} \bigl(\e_{ac}\bs{\mcl{B}}_{ib}(p)+\e_{cb}\bs{\mcl{B}}_{ ia}(p)\bigr)\mcl{Z}_{\b}\ ,\\
    \{\bs{G}_{\a a},\bs{\wt G}_{\b\wt a}\}\bs{\wt{\mcl{B}}}_{\wt{\imath}\wt c}(\wt p)|_{\bs{\Phi}_{b\wt b}}&=-\d_{\a\b}\e_{ a  b}e^{\a i\pi \wt p}\frac{\eta_{\wt p}^2}{x_{\wt p}} \bigl(\e_{\wt a\wt c}\bs{\mcl{\wt B}}_{\wt{\imath}\wt b}(\wt p)+\e_{\wt c\wt b}\bs{\mcl{\wt B}}_{ \wt{\imath}\wt a}(\wt p) \bigr)\mcl{Z}_{\a}\ ,\\
    \{\bs{G}_{\a a},\bs{\wt G}_{\b\wt a}\}\bs{\Psi}_{\c i}(p)|_{\bs{\Phi}_{b\wt b}}&=\e_{ab}\e_{\wt a \wt b}e^{-\b i\pi p}\frac{\eta_p^2}{x_p} \bigl(\d_{\b\c}\bs{\Psi}_{\a i}(p)+\e_{\a\c}\e_{\b\d}\bs{\Psi}_{ \d i}(p)\bigr)\mcl{Z}_{\b}\ ,\\
    \{\bs{G}_{\a a},\bs{\wt G}_{\b\wt a}\}\bs{\wt{\Psi}}_{\c\wt{\imath}}(\wt p)|_{\bs{\Phi}_{b\wt b}}&=-\e_{ a  b}\e_{\wt a \wt b}e^{\a i\pi \wt p}\frac{\eta_{\wt p}^2}{x_{\wt p}} \bigl(\d_{\a\c}\bs{\wt \Psi}_{\b\wt{\imath}}(\wt p)+\e_{\b\c}\e_{\a\d}\bs{\wt \Psi}_{\d\wt{\imath}}(\wt p)\bigr)\mcl{Z}_{\a}\ .
\end{aligned}
\ee
On a general multi-magnon state we then have to consider suitable sums of these expressions. For example, on a bosonic two-magnon state we find 
\begin{align*}
    \{\bs{G}_{\a a},\bs{\wt G}_{\b\wt a}\}\ket{\bs{\mcl{B}}_{i_1c_1}(p_1),\bs{{\mcl{B}}}_{i_2c_2}( p_2)}_{\bs{\Phi}_{b\wt b}}&\\=\d_{\a\b}\e_{\wt a \wt b} \Big[e^{-\b\cdot 2\pi i  p_2}e^{-\b i\pi p_1}\frac{\eta_{p_1}^2}{x_{p_1}}& \big| \bigl(\e_{ac_1}\bs{\mcl{B}}_{i_1b}(p_1)+\e_{c_1b}\bs{\mcl{B}}_{ i_1a}(p_1)\bigr),\bs{{\mcl{B}}}_{i_2c_2}( p_2)\mcl{Z}_{\b}\, \big\rangle\\ +e^{-\b i\pi p_2}\frac{\eta_{p_2}^2}{x_{p_2}}&\big| \bs{\mcl{B}}_{i_1c_1}(p_1),\bigl(\e_{ a c_2}\bs{\mcl{ B}}_{ i_2b}( p_2)+\e_{ c_2b}\bs{\mcl{ B}}_{ i_2a}( p_2)\bigr)\mcl{Z}_{\b} \, \big\rangle \Big]\ .
\end{align*}
It is easy to see that the expression vanishes if $a=b$, and so we may assume that $a\neq b$. In this case each line has only at most one non-zero term, and we find that they always cancel against one another, using that on physical states (for which $p_1+p_2\in\mbb{Z}$) we have the identity 
\be\label{phys1}
  e^{-\b\cdot 2\pi i  p_2}e^{-\b i\pi p_1}\frac{\eta_{p_1}^2}{x_{p_1}}+e^{-\b i\pi p_2}\frac{\eta_{p_2}^2}{x_{p_2}}=0\ .
\ee
If instead the second boson is a right-moving boson we find 
\begin{align*}
    \{\bs{G}_{\a a},\bs{\wt G}_{\b\wt a}\}\ket{\bs{\mcl{B}}_{ic}(p),\bs{\wt{\mcl{B}}}_{\wt{\imath}\wt c}(\wt p)}_{\bs{\Phi}_{b\wt b}}&\\=\d_{\a\b} \Big[\e_{\wt a \wt b}e^{\a\cdot 2\pi i \wt p}e^{-\a i\pi p}\frac{\eta_p^2}{x_p}&\big| \bigl(\e_{ac}\bs{\mcl{B}}_{ib}(p)+\e_{cb}\bs{\mcl{B}}_{ ia}(p)\bigr),\bs{\wt{\mcl{B}}}_{\wt{\imath}\wt c}(\wt p)\mcl{Z}_{\a}\, \big\rangle\\ -\e_{ a  b}e^{\a i\pi \wt p}\frac{\eta_{\wt p}^2}{x_{\wt p}}&\big| \bs{\mcl{B}}_{ic}(p), \bigl(\e_{\wt a\wt c}\bs{\mcl{\wt B}}_{\wt{\imath}\wt b}(\wt p)+\e_{\wt c\wt b}\bs{\mcl{\wt B}}_{ \wt{\imath}\wt a}(\wt p)\bigr)\mcl{Z}_{\a}\, \big\rangle \Bigr]\ .
\end{align*}
This vanishes by similar arguments as above, except that instead of (\ref{phys1}) we now need to employ 
\begin{equation}\label{phys2} 
e^{\a\cdot 2\pi i \wt p }e^{-\a i\pi p}\frac{\eta_p^2}{x_p}-e^{\a i\pi \wt p}\frac{\eta_{\wt p}^2}{x_{\wt p}}=0\ ,
\end{equation}
using that physical states respect $p-\tilde p\in \mbb{Z}$.
The other $2$-magnon cases work similarly, and for more general magnon excitations the argument can be generalised by making use of the telescoping identity 
\be\label{telescope}
\sum_{k=1}^n(1-e^{-2\pi i p_k}) \prod_ {j=k+1}^n e^{-2\pi i p_j} = (1-e^{-2\pi i \sum_k p_k})\ .
\ee
These arguments therefore work very similarly to those in \cite{Gaberdiel:2023lco}, which in turn were very reminiscent of the original ${\cal N}=4$ analysis of \cite{Beisert:2005tm}. It is straightforward to see that we can similarly define an $S$-matrix also in this general case, and that it will again satisfy the Yang-Baxter equation as in \cite{Gaberdiel:2023lco} --- this follows more or less directly from the fact that the commutators and anti-commutators behave covariantly with respect to the different $\mathfrak{su}(2)$ symmetries. In particular, our analysis therefore shows that integrability continues to be preserved also under the triplet perturbation. 

\subsection{Anomalous conformal dimensions}

Next we want to determine the (anomalous) conformal dimensions of arbitrary magnon descendants, and this works essentially as in \cite{Gaberdiel:2023lco}. In order to illustrate how this can be done, let us consider for simplicity a state that is only excited by right-moving magnons.\footnote{The general case can be analysed similarly, see \cite[eq.~(5.30) ff]{Gaberdiel:2023lco} for more details.} We can determine its anomalous conformal dimension by evaluating the action of 
\be
\{ {Q}_1, {S}_1 \}  = \{ {Q}_2, {S}_2 \}  = {\cal C} = ({L}_0 - {K}^3_0) \ ,
\ee
see eq.~(\ref{anti-com}), on it. This can be read off directly from the relations (\ref{3.2}), using again that, because of the fermionic statistics, only those terms contribute where both supercharges act on the same magnon, see the comment below eq.~(\ref{3.6}). However, in general, the image under, say, $\{ {Q}_1, {S}_1 \} $ will not be proportional to the original state, and in order to determine the anomalous conformal dimension we need to diagonalise the corresponding `mixing matrix'. 

For the case of the singlet perturbation, this mixing matrix is automatically diagonalisable, and hence the anomalous dimensions can be read off in this manner. However, for a general triplet perturbation, the mixing matrix will only be diagonalisable (with real eigenvalues) provided that the perturbation is real, and this is not automatic. In fact, in the basis we have considered above, the conjugate to the perturbation $\bs{\Phi}_{a\tilde{a}}$ is 
\be
\bs{\Phi}_{a\tilde{a}}^\dagger = \epsilon_{ab}\,  \epsilon_{\tilde{a} \tilde{b}}\, \bs{\Phi}_{b\tilde{b}} \ , \qquad \hbox{resp.} \quad 
\bs{\Phi}^\dagger_{a\wt{a}}=\frac{i}{\sqrt{2}}\begin{pmatrix}
        G_{\frac{1}{2}}^-\wt{G}_{\frac{1}{2}}^- & -G_{\frac{1}{2}}^-\wt{G}_{\frac{1}{2}}'^- \\
        -G_{\frac{1}{2}}'^-\wt{G}_{\frac{1}{2}}^- & G_{\frac{1}{2}}'^-\wt{G}_{\frac{1}{2}}'^-
    \end{pmatrix}|\text{BPS}_- \rangle_2\ ,
\ee
i.e.\ none of the individual perturbing fields are real, and we should instead consider the real combinations 
\be\label{real+img}
\Phi^{({\rm real})}_{a\tilde{a}} \equiv \bs{\Phi}_{a\tilde{a}} + \bs{\Phi}_{a\tilde{a}}^\dagger \ , \qquad 
\hbox{resp.} \qquad  
\Phi^{({\rm img})}_{a\tilde{a}} \equiv i \bigl(\bs{\Phi}_{a\tilde{a}} - \bs{\Phi}_{a\tilde{a}}^\dagger \bigr) \ . 
\ee
Thus instead of (\ref{3.2}) we need to evaluate the modified action of the supercharges under the real or imaginary perturbations of (\ref{real+img}); this leads to 
\begin{subequations}\label{3.2real}
\begin{align}
 \relax[\bs{G}_{\a a}, \bs{\wt{\mcl{B}}}_{\wt{\imath}\wt{a}}(\wt{p})]_{\bs{\Phi}^{({\rm real})}_{b\wt{b}}}&= -\e_{\a\b} \bigl( \e_{ab}\e_{\wt{a}\wt{b}} + \delta_{ab} \delta_{\tilde{a} \tilde{b}} \bigr) e^{\a i\pi \wt{p}}\frac{\eta_{\wt{p}}}{x_{\wt{p}}}\bs{\wt{\Psi}}_{\b\wt{\imath}}(\wt{p})\mcl{Z}_{\a}\ , \\
\{\bs{G}_{\a a}, \bs{\wt{\Psi}}_{\b\wt{\imath}}(\wt p)\}_{\bs{\Phi}^{({\rm real})}_{b\wt{b}}}&= \d_{\a\b}
\bigl( \e_{ab}  \delta_{\tilde{a}\tilde{b}}- \delta_{ab} \epsilon_{\tilde{a} \tilde{b}} \bigr) e^{\a i\pi \wt p}\frac{\eta_{\wt p}}{x_{\wt p}}\bs{\wt{\mcl{B}}}_{\wt{\imath}\wt{a}}(\wt p)\mcl{Z}_{\a}\ ,
\end{align}
\end{subequations}
and similarly 
\begin{subequations}\label{3.2img}
\begin{align}
 \relax[\bs{G}_{\a a}, \bs{\wt{\mcl{B}}}_{\wt{\imath}\wt{a}}(\wt{p})]_{\bs{\Phi}^{({\rm img})}_{b\wt{b}}}&= -i \e_{\a\b} \bigl( \e_{ab}\e_{\wt{a}\wt{b}} - \delta_{ab} \delta_{\tilde{a} \tilde{b}} \bigr) e^{\a i\pi \wt{p}}\frac{\eta_{\wt{p}}}{x_{\wt{p}}}\bs{\wt{\Psi}}_{\b\wt{\imath}}(\wt{p})\mcl{Z}_{\a}\ , \\
\{\bs{G}_{\a a}, \bs{\wt{\Psi}}_{\b\wt{\imath}}(\wt p)\}_{\bs{\Phi}^{({\rm img})}_{b\wt{b}}}&= i \d_{\a\b}
\bigl( \e_{ab}  \delta_{\tilde{a}\tilde{b}} + \delta_{ab} \epsilon_{\tilde{a} \tilde{b}} \bigr) e^{\a i\pi \wt p}\frac{\eta_{\wt p}}{x_{\wt p}}\bs{\wt{\mcl{B}}}_{\wt{\imath}\wt{a}}(\wt p)\mcl{Z}_{\a}\ .
\end{align}
\end{subequations}
In terms of the supercharges from above, see eq.~(\ref{superc}), these relations become 
\begin{subequations}
    \begin{align}
         \relax[S_1, \bs{\wt{\mcl{B}}}_{\wt{\imath}\wt{a}}(\wt{p})]_{\bs{\Phi}^{({\rm real})}_{b\wt{b}}}&= - \bigl( \e_{+b}\e_{\wt{a}\wt{b}} + \delta_{+b} \delta_{\tilde{a} \tilde{b}} \bigr) e^{ i\pi \wt{p}}\frac{\eta_{\wt{p}}}{x_{\wt{p}}}\bs{\wt{\Psi}}_{-\wt{\imath}}(\wt{p})\mcl{Z}_{+}\ , \\
         \{Q_1, \bs{\wt{\Psi}}_{-\wt{\imath}}(\wt p)\}_{\bs{\Phi}^{({\rm real})}_{b\wt{b}}}&= 
         \bigl( \e_{-b}  \delta_{\tilde{a}\tilde{b}}- \delta_{-b} \epsilon_{\tilde{a} \tilde{b}} \bigr) e^{- i\pi \wt p}\frac{\eta_{\wt p}}{x_{\wt p}}\bs{\wt{\mcl{B}}}_{\wt{\imath}\wt{a}}(\wt p)\mcl{Z}_{-}\ , \\
         \relax[Q_1, \bs{\wt{\mcl{B}}}_{\wt{\imath}\wt{a}}(\wt{p})]_{\bs{\Phi}^{({\rm real})}_{b\wt{b}}}&=  \bigl( \e_{-b}\e_{\wt{a}\wt{b}} + \delta_{-b} \delta_{\tilde{a} \tilde{b}} \bigr) e^{ -i\pi \wt{p}}\frac{\eta_{\wt{p}}}{x_{\wt{p}}}\bs{\wt{\Psi}}_{+\wt{\imath}}(\wt{p})\mcl{Z}_{-}\ , \\
         \{S_1, \bs{\wt{\Psi}}_{+\wt{\imath}}(\wt p)\}_{\bs{\Phi}^{({\rm real})}_{b\wt{b}}}&= 
         \bigl( \e_{+b}  \delta_{\tilde{a}\tilde{b}}- \delta_{+b} \epsilon_{\tilde{a} \tilde{b}} \bigr) e^{+ i\pi \wt p}\frac{\eta_{\wt p}}{x_{\wt p}}\bs{\wt{\mcl{B}}}_{\wt{\imath}\wt{a}}(\wt p)\mcl{Z}_{+}\ .
    \end{align}
\end{subequations}
Using the identity
\begin{equation}
    - \bigl( \e_{+b}\e_{\wt{a}\wt{b}} + \delta_{+b} \delta_{\tilde{a} \tilde{b}} \bigr)\bigl( \e_{-b}  \delta_{\tilde{a}\tilde{b}}- \delta_{-b} \epsilon_{\tilde{a} \tilde{b}} \bigr)+\bigl( \e_{-b}\e_{\wt{a}\wt{b}} + \delta_{-b} \delta_{\tilde{a} \tilde{b}} \bigr)\bigl( \e_{+b}  \delta_{\tilde{a}\tilde{b}}- \delta_{+b} \epsilon_{\tilde{a} \tilde{b}} \bigr)=1\ ,
\end{equation}
which follows from $\e_{\wt{a}\wt{b}} \delta_{\tilde{a} \tilde{b}}=0$ and $\delta_{\tilde{a} \tilde{b}}+(\e_{\wt{a}\wt{b}} )^2=1$, we then find that 
\begin{align}
    (L_0-K^3_0)\ket{\wt{\bs{\mcl{B}}}_{ \wt{\imath}\wt a}(\wt p)}_{\bs{\Phi}^{({\rm real})}_{b\wt{b}}}& =
    (L_0-K^3_0)\ket{\wt{\bs{\mcl{B}}}_{ \wt{\imath}\wt a}(\wt p)}_{\bs{\Phi}^{({\rm img})}_{b\wt{b}}} = 
    \left(\frac{\eta_{\wt p}}{x_{\wt p}}\right)^2\ket{\wt{\bs{\mcl{B}}}_{ \wt{\imath}\wt a}(\wt p)}\ , \\
    (L_0-K^3_0)\ket{\wt{\bs{\Psi}}_{\a \wt{\imath} }(\wt p)}_{\bs{\Phi}^{({\rm real})}_{b\wt{b}}}&=
    (L_0-K^3_0)\ket{\wt{\bs{\Psi}}_{\a \wt{\imath} }(\wt p)}_{\bs{\Phi}^{({\rm img})}_{b\wt{b}}} = \left(\frac{\eta_{\wt p}}{x_{\wt p}}\right)^2\ket{\wt{{\Psi}}_{ \a \wt{\imath}}(\wt p)}\ ,
\end{align}
which therefore has the same form as eq.~(\ref{singletano}). In particular, it therefore follows that all the individual magnon modes have the same dispersion relation of eq.~(\ref{disp}) for the triplet perturbation as in the singlet case.

\subsection{Long magnon modes under the triplet perturbation}\label{long}

The above magnons correspond to the torus excitations in the dual AdS background \cite{Frolov:2023pjw}, but as was realised in \cite{Gaberdiel:2024dfw}, there are new collective magnons (dubbed `long magnons' in \cite{Gaberdiel:2024dfw}) that appear in the large $w$ limit, and that account for the ${\rm AdS}_3 \times {\rm S}^3$ excitations of the string background. These collective modes are bilinear combinations of the torus modes --- they are small deformations of the fractional ${\cal N}=4$ modes \cite{Lunin:2002fw,Gomis:2002qi,Gava:2002xb} --- but since the sum involves ${\cal O}(w)$ terms, they need to be treated as independent modes at large $w$. 

Given that the dispersion relation for these long magnons cannot be directly read off from that of the  individual magnons above, we also need to study how they behave with respect to the triplet perturbation. Fortunately, we do not have to redo the entire analysis since we can read off most of what we are interested in using symmetry arguments. As we have explained above, see eq.~(\ref{3.1}), the perturbing fields transform in the $({\bf 2},{\bf 2})$ with respect to $\mfr{su}(2)_{[1]}\oplus \tilde{\mfr{su}}(2)_{[1]}$, and the relations in (\ref{3.2}) transform covariantly with respect to these symmetries.\footnote{Strictly speaking, these are the relations in the large $w$ limit, while for the analysis of the long magnons we can only take the large limit after we have evaluated the individual matrix elements, i.e.\ we need the matrix elements $b_{n}^{m}$ and $c_{n}^{m}$ in eq.~(\ref{deformedaction}) at finite $w$ before taking the large $w$ limit. However, the covariant structure of these relations is also true at finite $w$.} Thus in order to see how the singlet deformation results for the long magnons are modified, we only need to study how they transform with respect to $\mfr{su}(2)_{[1]}\oplus \tilde{\mfr{su}}(2)_{[1]}$. Since this transformation only affects the bosons, it is clear that the $\Xi$ magnon of \cite[eq.~(2.20)]{Gaberdiel:2024dfw} that consists of fermion bilinears is invariant under these transformations, and hence behaves exactly the same under the triplet deformation. The same turns out to be true for the $\Lambda$ magnon of \cite[eq.~(3.9)]{Gaberdiel:2024dfw} that has the schematic form 
\ba
\Lambda_{-1+\frac{n}{w}}
   &= \tfrac{1}{\sqrt{(w+1-n)(1-\frac{n}{w})^2}}\Bigg(\sum_{a=n+1}^{w-1}\xi(\tfrac{n}{w};\tfrac{a}{w})\:\Big[\bar{\a}^1_{\frac{n}{w}-\frac{a}{w}}\a^2_{-1+\frac{a}{w}}
    -\a^1_{\frac{n}{w}-\frac{a}{w}}\bar{\a}^2_{-1+\frac{a}{w}}\Big] \nonumber \\
    & \quad +\sum_{a=n+1}^w \xi(\tfrac{n}{w};\tfrac{a}{w}) \: (\tfrac{a}{w}-\tfrac{n}{w})\Big[\bar{\psi}^+_{-\frac{1}{2}+\frac{n}{w}-\frac{a}{w}}\psi^-_{-\frac{1}{2}+\frac{a}{w}}
    - \psi^+_{-\frac{1}{2}+\frac{n}{w}-\frac{a}{w}}\bar{\psi}^-_{-\frac{1}{2}+\frac{a}{w}}\Big]\Bigg) \ .
\end{align}
In order to see this we first note that the $\xi(\tfrac{n}{w};\tfrac{a}{w})$ coefficients satisfy 
\be
    \xi(\tfrac{n}{w};\tfrac{a}{w})= \xi(\tfrac{n}{w};1+\tfrac{n}{w}-\tfrac{a}{w})\ .
\ee
Thus we can rewrite the first line --- it is enough to consider the first line since the fermions are unaffected by the bosonic rotation by $\mfr{su}(2)_{[1]}$ --- as 
\be
\sum_{a=n+1}^{w-1}\xi(\tfrac{n}{w};\tfrac{a}{w})\:\Big[\bar{\a}^1_{\frac{n}{w}-\frac{a}{w}}\a^2_{-1+\frac{a}{w}}
    -\a^1_{-1+\frac{a}{w}}\bar{\a}^2_{\frac{n}{w}-\frac{a}{w}}\Big] \ . 
\ee
Using eq.~(\ref{genbos}), it is then easy to see that the combination in brackets is invariant under an arbitrary transformation by a group element in $\mfr{su}(2)_{[1]}$. According to the dictionary in \cite[eq.~(5.6)]{Gaberdiel:2024dfw}, this then means that the bosonic magnons associated to the ${\rm AdS}_3 \times {\rm S}^3$ directions behave the same way under the triplet vs.\ the singlet perturbation. This suggests that, from a spacetime perspective, the perturbation modifies only the $\mbb{T}^4$ directions, see also the discussion in Section~\ref{sec: supergravity}. 
\smallskip

\noindent On the other hand, the fermionic generators, see \cite[eq.~(3.9)]{Gaberdiel:2024dfw}
\ba\label{eq:other_bilinear_magnons}
\Gamma^-_{-\frac{1}{2}+\frac{n}{w}}
&= \tfrac{1}{\sqrt{(w+1-n)(1-\frac{n}{w})}}\sum_{a=n+1}^w\xi(\tfrac{n}{w};\tfrac{a}{w})\Big[\a^1_{\frac{n}{w}-\frac{a}{w}}\bar{\psi}^-_{-\frac{1}{2}+\frac{a}{w}}
+\bar{\a}^1_{\frac{n}{w}-\frac{a}{w}}\psi^-_{-\frac{1}{2}+\frac{a}{w}}\Big]
\nonumber \\
\Gamma'^-_{-\frac{1}{2}+\frac{n}{w}}
&= \tfrac{1}{\sqrt{(w+1-n)(1-\frac{n}{w})}}\sum_{a=n+1}^w\xi(\tfrac{n}{w};\tfrac{a}{w})\Big[\a^2_{\frac{n}{w}-\frac{a}{w}}\bar{\psi}^-_{-\frac{1}{2}+\frac{a}{w}}
+\bar{\a}^2_{\frac{n}{w}-\frac{a}{w}}\psi^-_{-\frac{1}{2}+\frac{a}{w}}\Big]
\ , \nonumber
\end{align}
transform under $\mfr{su}(2)_{[1]}$ as a doublet, i.e.\ as 
\be
\begin{pmatrix} a\ \ & b \\ c\ \ & d \end{pmatrix}\in \mfr{su}(2)_{[1]} \qquad \qquad \hbox{as} \qquad 
\Biggl( \begin{array}{c} \Gamma'^-_{-\frac{1}{2}+\frac{n}{w}}  \\
 \Gamma^-_{-\frac{1}{2}+\frac{n}{w}} \end{array} \Biggr) \quad \mapsto \quad 
\begin{pmatrix} a\ \ & c \\ b\ \ & d \end{pmatrix} \,  
\Biggl( \begin{array}{c} {\displaystyle \Gamma'^-_{-\frac{1}{2}+\frac{n}{w}}}  \\
{\displaystyle \Gamma^-_{-\frac{1}{2}+\frac{n}{w}}} \end{array} \Biggr)  \ ,
\ee
i.e.\ the same way as $(G'^{-},G^-)$, see eq.~(\ref{GtildeG}). This is to be expected, because contrary to the bosons, the fermions associated to AdS$_3\times $S$^3$ do not decouple from those associated to the $\mbb{T}^4$. This is as in flat space: the bosons sit in the vector representation ${\bf 8}_v$ of $\mathfrak{so}(8)$ that decomposes as ${\bf 8}_v = ({\bf 4},{\bf 1}) \oplus ({\bf 1},{\bf 4})$ with respect to $\mathfrak{so}(4) \oplus \mathfrak{so}(4) \subset \mathfrak{so}(8)$, i.e.\ the bosons from the different directions decouple from one another. On the other hand, the spacetime fermions sit in a spinor representation, say ${\bf 8}_s$ of $\mathfrak{so}(8)$, and that decomposes as ${\bf 8}_s =({\bf 2}_s,{\bf 2}_s) \oplus ({\bf 2}_c,{\bf 2}_c)$, i.e.\ as a bispinor.

\section{Type IIB supergravity \label{sec: supergravity}}

In this chapter we want to identify the different perturbations of the symmetric orbifold with deformations of type IIB supergravity on AdS$_3\times $S$^3\times \mbb{T}^4$. While the symmetric orbifold point itself does not have a direct supergravity interpretation --- it describes the pure NS-NS background with minimal flux  \cite{Gaberdiel:2018rqv,Eberhardt:2018ouy,Eberhardt:2019ywk} --- we can still use the structure (and the symmetries) of supergravity with large NS-NS flux to get a good sense of what our CFT perturbations are dual to. 
Our analysis will be somewhat similar to what was already reviewed in \cite{David:2002wn}. However, at the time there was some confusion about the spacetime dual of the symmetric orbifold, and hence some of the conclusions of \cite{David:2002wn} will have to be modified.\footnote{A careful analysis of the moduli space of these backgrounds, taking these recent advances into account, was done in \cite{Aharony:2024fid}. However, they only concentrated on the `singlet' part of the moduli space.}

We will begin by reviewing the supergravity solutions that are relevant for our set-up. We will then show  that their deformations exhibit the same symmetries as those present in the dual CFT, and this will allow us to identify the supergravity solutions that are switched on by the triplet perturbation of the symmetric orbifold.

\subsection{Conventions and the class of solution} 

The bosonic fields of ten-dimensional type IIB supergravity consist of the ten-dimensional (Lorentzian) metric $g$, the dilaton $\phi$, the axion or R-R scalar $C^{(0)}$, the R-R gauge potentials $C^{(2)}$ and $C^{(4)}$, and the NS-NS gauge potential $B^{(2)}$.\footnote{In the following we shall be using the conventions of \cite{Figueroa-OFarrill:2012whx,Wulff:2017zbl}.} The metric is a symmetric rank two tensor, with signature corresponding to one timelike direction and nine spacelike directions. The dilaton is a scalar field which is linked through its vacuum expectation value to the string coupling constant, $g_s=e^{\langle\phi\rangle}$. The R-R scalar is a differential zero-form, or equivalently a scalar field. The R-R gauge potentials $C^{(2)}$ and $C^{(4)}$ are a 2-form and a 4-form, respectively, while the NS-NS gauge potential $B^{(2)}$ is a 2-form. The corresponding field strengths are then \begin{equation}
  \begin{aligned}
    G^{(1)} &= d C^{(0)}\\
    G^{(3)} &= d C^{(2)} - C^{(0)} \wedge H^{(3)}\\
    H^{(3)} &= d B^{(2)}\\
    G^{(5)} &= d C^{(4)} - \frac{1}{2} d B^{(2)} \wedge C^{(2)} +
    \frac{1}{2} d C^{(2)} \wedge B^{(2)}~.
  \end{aligned}
\end{equation}
In the near-horizon geometry of the NS5-F1 brane system --- this is the brane system that generates AdS$_3$ solutions with pure NS-NS flux --- some of the moduli are fixed by the attractor mechanism, and the remaining moduli are the $16$ torus moduli, as well as the R-R scalar $C^{(0)}$ and the self-dual R-R 2-form \cite{Larsen:1999uk,OhlssonSax:2018hgc}. The relevant solutions are of the form, see e.g.\ \cite{Figueroa-OFarrill:2012whx,Wulff:2017zbl,OhlssonSax:2018hgc}
\begin{equation}   
H^{(3)}= f_{\rm NS} \,(\nu+\sigma)\ ,\hspace{1cm} G^{(5)}=   (\nu+\sigma) \wedge \omega\ ,\hspace{1cm} G^{(3)}=f_{\rm R}\, (\nu+\sigma)\ ,
\end{equation}
where $\nu$ and $\sigma$ are normalised 3-forms on ${\rm AdS}_3$ and ${\rm S}^3$, respectively, while $\omega$ is a self-dual 2-form on $\mathbb{T}^4$. A simple basis of normalised $2$-forms on $\mathbb{T}^4$ is given by 
\begin{align}
    & \omega^1_\pm=\frac{1}{\sqrt{2}} \bigl( dx^1\wedge dx^4\pm dx^2\wedge dx^3 \bigr)\\
    & \omega^2_\pm =\frac{1}{\sqrt{2}} \bigl(  dx^1\wedge dx^3\pm dx^4\wedge dx^2 \bigr)\\
    & \omega^3_\pm =\frac{1}{\sqrt{2}} \bigl(  dx^1\wedge dx^2\pm dx^3\wedge dx^4 \bigr)\  ,
\end{align}
where $\omega^i_+$ with $i=1,2,3$ are self-dual, while $\omega^i_-$ are anti-self-dual. We can thus expand 
\be\omega = c_i \, \omega^i_+ \ .
\ee
With these conventions, the radius of the ${\rm AdS}_3$ and ${\rm S}^3$ are both equal to 
\be
R =  \sqrt{f_{\rm NS}^2 + f_{\rm R}^2 + \vec{c}^{\,\, 2}} \ . 
\ee
Starting from the background with $f_{\rm NS}=1$, $f_{\rm R}=\vec{c}=0$ --- this is the background that is dual to the symmetric orbifold --- we can either switch on $f_{\rm R}$ or $\vec{c}$. Both affect equally the ${\rm AdS}_3$ and ${\rm S}^3$ geometry, but they have a different effect on the $\mathbb{T}^4$ excitations. We now want to argue that switching on $f_{\rm R}$ corresponds to the singlet perturbation of Section~\ref{singletpert} that was studied in detail in \cite{Gaberdiel:2023lco}. On the other hand, switching on $\vec{c}$ describes the triplet perturbation we have studied in Section~\ref{tripletpert}. 

As a first piece of evidence we note that both perturbations have the same effect on the bosonic long magnon modes associated to the ${\rm AdS}_3$ and ${\rm S}^3$ directions, see Section~\ref{long}, and this is in agreement with the above supergravity analysis: the radius, resp.\ the curvature, of either space depends only on the combination $f_{\rm R}^2 + \vec{c}^{\,\, 2}$, so switching on $f_{\rm R}$ or $\vec{c}$ has the same effect on it. 
On the other hand, the action of the triplet perturbation on the torus modes is different from that of the singlet perturbation, and this is also obvious from the above supergravity analysis (since choosing a specific vector $\vec{c}$ breaks the full rotational symmetry of the problem). 

To be more specific, in the symmetric orbifold the two classes of perturbations are distinguished by their transformation property with respect to $\mathfrak{su}(2)_{[1]}$. This also has a very natural counterpart in the above supergravity description. To see this we recall that there is a natural $\mathfrak{so}(4)  = \mathfrak{su}(2)_{\ell} \oplus \mathfrak{su}(2)_{r}$ action on the 4 bosons, where the two $\mathfrak{su}(2)$'s act from the left and from the right on the matrix 
\be
\tilde{x} = \begin{pmatrix} x_1 + i x_2 & - x_3 + i x_4 \\  x_3 + i x_4 & x_1 - i x_2 \end{pmatrix} \qquad \hbox{with determinant}\  \det(\tilde{x}) = \sum_{i=1}^{4} x_i^2 \ . 
\ee
Indeed the map $\tilde{x} \mapsto g_{\ell} \cdot \tilde{x} \cdot g_{r}$ with $g_{\ell/r}\in {\rm SU}(2)$ preserves the norm of the 4-vector $x_i$, and hence defines a rotation of $x_i$. Given this explicit $ \mathfrak{su}(2)_{\ell} \oplus \mathfrak{su}(2)_{r}$ action on the $4$ bosons, one can easily determine how the self-dual and anti-self-dual $2$-forms transform, and we find that the self-dual $2$-forms sit in a triplet of $\mathfrak{su}(2)_{r}$ (and are invariant with respect to $\mathfrak{su}(2)_{\ell}$), while the roles of $\mathfrak{su}(2)_{r}$ and $\mathfrak{su}(2)_{\ell}$ are interchanged for the anti-self-dual 2-forms. 

By comparing the above action on the bosons with eq.~(\ref{genbos}),  it is natural to identify (recall that $\mathfrak{su}(2)_{\rm B}^{(1)} = \mathfrak{su}(2)_{[1]}$)
\be
\mathfrak{su}(2)_{\ell} \oplus \mathfrak{su}(2)_{r} \cong \mathfrak{su}(2)_{\rm B}^{(2)} \oplus \mathfrak{su}(2)_{\rm B}^{(1)}  \ .
\ee
Thus the dual of the triplet perturbation of the symmetric orbifold should be a family of solutions that transforms as a triplet with respect to $\mathfrak{su}(2)_{r}$, and this is exactly realised by the supergravity deformations labelled by $\vec{c}$. We regard this as rather convincing evidence that the triplet perturbation of the symmetric orbifold corresponds to switching on the $\vec{c}$ supergravity parameters. 

As mentioned, this is very similar to what was proposed in \cite{David:2002wn}, except that now it is the self-dual R-R 2-forms that are switched on --- in \cite{David:2002wn} it was instead claimed that the relevant supergravity deformation involves the self-dual NS-NS 2-forms, see \cite[eq.~(6.16)]{David:2002wn}. (The reason for this mismatch is that, at the time, it was believed that the dual of the symmetric orbifold has pure R-R flux.)

\section{Conclusion \label{sec: conclusion}}    

In this paper we have studied the triplet perturbation of the symmetric orbifold, using similar techniques as those employed in \cite{Gaberdiel:2023lco}. As we have seen, the structure of the deformed algebra is very similar to what happened in the singlet case, and as a consequence the triplet perturbation is also integrable. Using symmetry arguments, as well as the fact that the triplet and single perturbation behave the same way with respect to the bosonic ${\rm AdS}_3 \times {\rm S}^3$ magnons that were identified in \cite{Gaberdiel:2024dfw}, we could identify the AdS deformation as being associated to switching on the self-dual part of the R-R 2-form. (This is obviously in line with what was to be expected, but it is reassuring to confirm this by a direct analysis.) 

While these perturbative considerations only explore the neighbourhood of the pure NS-NS point, it would be very interesting to understand the structure of the theory (and in particular its spectrum) near the point where R-R flux dominates. In particular, one may hope in this way to make contact with the predictions of, e.g.\ \cite{Cavaglia:2021eqr,Cavaglia:2022xld}. 

\section*{Acknowledgements}

This paper is largely based on the Master thesis of one of us (ILeM). We thank Beat Nairz for helpful conversations and useful comments on a draft version of this paper.  The work of the group at ETH is supported by a personal grant of MRG from the Swiss National Science Foundation, by the Simons Foundation grant 994306 (Simons Collaboration on Confinement and QCD Strings), as well as the NCCR SwissMAP that is also funded by the Swiss National Science Foundation.

\appendix

\section{Conventions}\label{app:conventions}

\begingroup
\allowdisplaybreaks

In this appendix we collect our conventions for the torus fields and the ${\cal N}=4$ superconformal generators.

\subsection{$\mcl{N}=4$ algebra}\label{app:algebra}

We denote the left-moving four bosons and four fermions of the $\mbb{T}^4$ by $\a^i,\bar{\a}^i$, $i=1,2$ and $\psi^\pm,\bar{\psi}^\pm$, respectively. They satisfy the OPE relations
\be
\bar{\a}^i(x)\a^j(y)\sim \frac{\epsilon^{ij}}{(x-y)^2}\ ,\qquad \bar{\psi}^\pm(x)\psi^\mp(y)\sim \frac{\pm 1}{x-y} \ ,
\ee
with $\epsilon^{12}=1$. Right-movers are always denoted by a tilde. The (left-moving) $\mcl{N}=4$ generators are built out of these fields as
\begin{equation}
    \bs{G}_{\a b}=\bs{{\Psi}}_{\a i}\,\e_{ij}\,\bs{{\mcl{B}}}_{j{b}}\ , \qquad K^a=\frac{1}{2}\e_{ij}\bs{{\Psi}}_{\a j}\e_{\a\b}D(t^a)_{\c\b}  \bs{{\Psi}}_{\c i}\ ,
\end{equation}
where $\bs{{\Psi}}_{\a i}$ and $\bs{{\mcl{B}}}_{j{b}}$ are defined in eqs.~(\ref{genferm}) and (\ref{genbos}), respectively, and $\alpha=\pm$ is a spinor index, with $\e_{+-}=-\e_{-+}=1$. Here $D(t^a)_{\b\c}$ are the matrix elements of the $\mfr{su}(2)$ generators in the $j=\frac{1}{2}$ representation, 
\be
D(t^3)= \left( \begin{matrix} \frac{1}{2} & 0 \cr 0 & -\frac{1}{2} \end{matrix} \right) \ , \qquad 
D(t^+)= \left( \begin{matrix} 0 & -1 \cr 0 & 0 \end{matrix} \right) \ , \qquad 
D(t^-)= \left( \begin{matrix} 0 & 0 \cr -1 & 0 \end{matrix} \right) \ . 
\ee
Finally, the stress energy tensor takes the usual free field form,
\be
L =  \,:\bar{\a}^1\,\a^2-\bar{\a}^2\,\a^1:+\,\frac{1}{2}:\psi^+\,\partial \bar{\psi}^- + \bar{\psi}^-\,\partial \psi^+ - \bar{\psi}^+\,\partial\psi^- - \psi^-\,\partial\bar{\psi}^+:\ .
\ee
These fields generate then the $c=6$, ${\cal N}=4$ superconformal algebra, whose modes satisfy
\be\label{N4rel}
\begin{array}{rcl}
{}   [L_m,L_n]&=& (m-n)L_{m+n}+\tfrac{1}{2}(m^3-m)\d_{m+n,0}\\
{}   [L_m,K^a_n]&= & -nK^a_{m+n}\\
{}   [L_m,(\bs{G}_{\a b})_r]&= & \big(\tfrac{m}{2}-r\big)(\bs{G}_{\a b})_{m+r}\\
{}   \{(\bs{G}_{\pm +})_r,(\bs{G}_{\mp -})_s\}&= & \pm L_{r+s}+(r-s)K^3_{r+s} \pm \big(r^2-\tfrac{1}{4}\big)\d_{r+s,0}\\
{}   \{(\bs{G}_{\pm +})_r,(\bs{G}_{\pm -})_s\}&= & \pm(r-s)K^\pm_{r+s}\ ,
\end{array}
\ee
as well as
\be
\begin{array}{cclccl}
{}[K^3_m,K^3_n] &=& \tfrac{m}{2}\d_{m+n,0} \qquad   & [K^3_m,(\bs{G}_{\pm b})_r]&=& \pm\tfrac{1}{2}(\bs{G}_{\pm b})_{m+r} \\
{} [K^3_m,K^\pm_n]&=& \pm K^\pm_{m+n}\qquad &[K^\pm_m,(\bs{G}_{\mp b})_r]&=& -(\bs{G}_{\pm b})_{m+r}\\
{}[K^+_m,K^-_n]&= & 2K^3_{m+n}+m\,\d_{m+n,0}\ .\qquad & & &
\end{array}
\ee
The other (anti-)commutators vanish. In order to make contact with the conventions of \cite{Gaberdiel:2023lco}, we can write the different components of $\bs{G}_{\a b}$ as in (\ref{GtildeG}); then the corresponding fields have the free field realisation
\be\label{N4fields}
\begin{array}{cclccl}
G^+ &= & \bar{\a}^2\,\psi^++\a^2\,\bar{\psi}^+ \qquad & K^+ &= & \bar{\psi}^+\,\psi^+ \\
G'^+ &= & -\bar{\a}^1\,\psi^+-\a^1\,\bar{\psi}^+\qquad & K^- &= & - \bar{\psi}^-\,\psi^- \\
G^- &= & \bar{\a}^1\,\psi^-+\a^1\,\bar{\psi}^- \qquad & K^3 &= & \frac{1}{2}\, \bigl( :\bar{\psi}^+\,\psi^-:+:\bar{\psi}^-\,\psi^+: \bigr) \\
G'^- &= & \bar{\a}^2\,\psi^-+\a^2\,\bar{\psi}^-\ . \qquad & &&
\end{array}
\ee

\endgroup

\subsection{States}\label{app:states}

In the symmetric product orbifold of $\mbb{T}^4$, we denote the $w$-twisted sector ground-state by $\sigma_w$. For odd $w$ it has conformal dimension $h=\wt{h}=\tfrac{w^2-1}{4w}$ and $\mfr{su}(2)$ charge $m=\wt{m} = 0$. For even $w$, there is actually a doublet of Ramond ground states, and we denote by $\sigma_w$ the state with $h=\wt{h}=\tfrac{w}{4}$ and $m=\wt{m}=-\tfrac{1}{2}$.

In each twisted sector, there are four (left and right) BPS states with charge and dimension $\frac{w-1}{2},\frac{w}{2},\frac{w+1}{2}$. We denote the BPS state with $h=\wt{h}=m=\wt{m}=\tfrac{w-1}{2}$ by $\ket{\text{BPS}_-}_w$. Explicitly, this state is
\ba
\ket{\mrm{BPS}_-}_w &= \big(\bar{\psi}^+_{-\frac{w-2}{2w}}\psi^+_{-\frac{w-2}{2w}}\cdots\bar{\psi}^+_{-\frac{1}{2w}}\psi^+_{-\frac{1}{2w}}\big)\big(\wt{\bar{\psi}}^+_{-\frac{w-2}{2w}}\wt{\psi}^+_{-\frac{w-2}{2w}}\cdots\wt{\bar{\psi}}^+_{-\frac{1}{2w}}\wt{\psi}^+_{-\frac{1}{2w}}\big)\sigma_w\ ,\quad\text{for $w$ odd}\ ,\nonumber \\
\ket{\mrm{BPS}_-}_w &= \big(\bar{\psi}^+_{-\frac{w-2}{2w}}\psi^+_{-\frac{w-2}{2w}}\cdots\bar{\psi}^+_{0}\psi^+_{0}\big)\big(\wt{\bar{\psi}}^+_{-\frac{w-2}{2w}}\wt{\psi}^+_{-\frac{w-2}{2w}}\cdots\wt{\bar{\psi}}^+_{0}\wt{\psi}^+_{0}\big)\sigma_w\ ,\quad\text{for $w$ even}\ .
\end{align}
The other BPS states can be obtained by applications of $\bar{\psi}^+_{-1/2}$, $\psi^+_{-1/2}$ and the corresponding right-movers. The standard BPS state we work with has $h=\wt{h}=m=\wt{m}=\frac{w+1}{2}$ and is given by
\be
\ket{\mrm{BPS}}_w ={\psi}^+_{-\frac{1}{2}}\bar{\psi}^+_{-\frac{1}{2}}\wt{{\psi}}{}^+_{-\frac{1}{2}}\wt{\bar{\psi}}^+_{-\frac{1}{2}}\ket{\mrm{BPS}_-}_w\ .
\ee
The perturbing fields are descendants of the lower BPS state in the two-twisted sector, and are given explicitly in eq.~(\ref{eq: four perturbations}). They then have $h=\wt{h}=1$ and $m=\wt{m}=0$.

\bibliographystyle{JHEP}  

\end{document}